\documentclass{PoS}
\usepackage[latin9]{inputenc}
\usepackage{units}
\usepackage{amsmath}
\makeatletter

\title{The perturbative SU(N) one-loop running coupling in the twisted gradient flow scheme}

\ShortTitle{The perturbative SU(N) one-loop running coupling in the TGF scheme}

\author{\speaker{Eduardo I. Bribián}, Margarita García Pérez%
\\
       Instituto de Física Teórica UAM-CSIC, Madrid\\
       Nicolás Cabrera 13-15, Universidad Autónoma de Madrid, E-28049-Madrid,Spain\\
       E-mail: \email{e.i.bribian@csic.es, margarita.garcia@uam.es}}

\abstract{We report on our computation of the perturbative running of the 't Hooft coupling in a pure gauge SU(N) theory with twisted boundary conditions. 
The coupling is defined in terms of the energy density of the flow fields at a scale given by a particular combination of the linear size of the 
torus and the rank of the gauge group. We present our results for the matching at one-loop order to the $\MS$ scheme in the case of a two-dimensional non-trivial twist. 
The ratio of $\Lambda$ parameters is determined for the case of $SU(3)$ and for various other values of the number of colours and 
several choices of the magnetic flux induced by the twist.}

\FullConference{36th annual International Symposium on Lattice Field Theory\\
22-28 July 2018\\
Michigan State University, East Lansing, USA}

\makeatother

\newcommand{\comment}[1]{}

\newcommand{\be}{\begin{equation}}
\newcommand{\ee}{\end{equation}}
\newcommand{\ba}{\begin{array}}
\newcommand{\ea}{\end{array}}
\newcommand{\baa}{\begin{array}}
\newcommand{\eaa}{\end{array}}
\newcommand{\bea}{\begin{eqnarray}}
\newcommand{\eea}{\end{eqnarray}}
\newcommand{\bal}{\begin{align}}
\newcommand{\eeal}{\end{align}}
\newcommand{\half}{\frac{1}{2}}

\newcommand{\Tr}{\mathrm{Tr}}
\newcommand{\lef}{\tilde l}

\newcommand{\hc}{{\hat c}}
\newcommand{\tl}{\tilde l}

\newcommand{\htheta}{\hat \theta}
\newcommand{\TGF}{{\rm TGF}}
\newcommand{\cE}{{\mathcal E}}

\newcommand{\rd}{{\rm div}}
\newcommand{\rf}{{\rm fin}}

\newcommand{\MS}{{\overline{\rm MS}}}
\newcommand{\cA}{{\mathcal A}}

\newcommand{\cC}{ \mathcal{C}}
\newcommand{\Phio}{ \Phi^{(0)}}

\begin{document}

\section{Introduction}

	The gradient flow~\cite{Narayanan:2006rf,Luscher:2009eq}, alongside finite-size scaling methods, has been extensively used to study the scale dependence of the gauge coupling constant. A variety of gradient-flow based renormalisation schemes have been introduced to this goal~\cite{Luscher:2010iy}-\cite{Ramos:2014kla}. Perturbative calculations in these schemes are, however, comparatively rare. In particular, next-to-next to leading order (NNLO) matching to $\MS$ has been achieved in infinite volume for the $SU(N)$ gauge group with $N_f$ flavours of massless quarks~\cite{Luscher:2010iy,Harlander:2016vzb}, while in finite volume it has been done, using numerical stochastic perturbation theory, for the $SU(3)$ pure gauge theory in the Schr\"odinger functional scheme~\cite{DallaBrida:2017tru}.

	In these proceedings, we will report on our next to leading order (NLO) perturbative calculation of the $SU(N)$ running coupling in the twisted gradient flow (TGF) scheme, a finite volume scheme introduced by A. Ramos in~\cite{Ramos:2014kla} and used recently for a precise non-perturbative determination of the $\Lambda_\MS$ parameter in $SU(3)$ pure gauge theory~\cite{Ishikawa:2017xam}. This set-up has also been used, in combination with the idea of twisted volume reduction~\cite{Eguchi:1982nm}-\cite{GonzalezArroyo:2010ss}, to determine the running of the $SU(\infty)$ coupling constant on a one-site lattice, using the rank of the gauge group to implement step scaling~\cite{GarciaPerez2015}. The calculation exploits the idea that in the large $N$ limit space-time and colour degrees of freedom become entangled to the point where the former are redundant. This is reflected in our finite volume prescription in the parametrisation of the running coupling in terms of an effective size $\tl$ that depends on both the physical size of the torus and the number of colours of the theory. The general strategy for performing the perturbative calculation has been described in~\cite{Bribian:2016qvf}. 

	The perturbative computation will be done in the continuum, and will consist of a few main steps: first, we will define the twisted gradient flow coupling. Then, we will expand the fields in perturbation theory, and rewrite the leading order (LO) and NLO contributions in terms of either Jacobi theta functions or a few integrals of Siegel theta functions. We will then explain the procedure we devised to regularise the UV divergent integrals in dimensional regularisation by computing and subtracting the asymptotic divergence. Lastly, we will relate our scheme to the $\MS$ one, computing the ratio of $\Lambda$ parameters, and present our numerical results.

\section{The twisted gradient flow scheme}

	We will begin by defining the scheme with which we will work. Consider a pure gauge $SU(N)$ theory on a $d$-dimensional torus with twisted boundary conditions (TBC). Denoting $l_\mu$ the length of each side, and $d_t\le d$ an even number, the gauge fields satisfy:
	\begin{align}
		& A_\mu ( x + l_\nu \hat \nu) = \Gamma_\nu A_\mu (x) \Gamma_\nu^\dagger, \quad  \nu= 0, \dots, d_t-1 \ , \\
		& A_\mu ( x + l_\nu \hat \nu) = A_\mu (x), \quad  \nu= d_t, \dots, d \ .
	\end{align}
	The $\Gamma_\mu$ matrices, known as twist-eaters, satisfy the consistency condition:
	\begin{equation}
		 \Gamma_\mu \Gamma_\nu  = \exp\{ i 2 \pi \epsilon_{\mu \nu} k /l_g \} \Gamma_\nu \Gamma_\mu , \quad
		l_g=N^{2/d_t} \, ,
	\end{equation}
	with  $k$ and $l_g$ taken to be two co-prime integers. We will henceforth set $d=4$ and take $d_t=2$, with  $\epsilon_{01}=-\epsilon_{10}=1$, and $\epsilon_{\mu\nu}=0$ elsewhere. The procedure can be extended to the case of $d_t=4$ with $\epsilon_{\mu \nu}$ antisymmetric and $\epsilon_{\mu \nu}=1$ for $\mu<\nu$, where the twist becomes non-trivial in all planes.
		
	Under such boundary conditions, one can Fourier-expand the gauge fields using a momentum-dependent basis of the $SU(N)$ Lie algebra $\hat{\Gamma}(q)$~\cite{GonzalezArroyo:1982hz,GonzalezArroyo:1983ac}, in terms of which:
	\begin{equation}
		A_\mu ( x) = V^{-\half} \underset{q}{\sum^{\prime}} \hat A_\mu (q) e^{i q x} \, \hat{\Gamma}(q)\, ,  
	\end{equation}
	where all matrix structure is contained in $\hat{\Gamma}(q)$, $V$ is the volume of the torus, and the prime denotes the exclusion of all momenta proportional to $l_g$. 
	Defining several auxiliary variables, $\theta_{\mu\nu}=l_g^2 l_\mu l_\nu \tilde \epsilon_{\mu \nu} \htheta/(2\pi)$, $\htheta  = \bar k/l_g $, $k\bar{k}=1 ({\rm mod}\, l_{g})$, and $\underset{\nu}{\sum}\tilde{\epsilon}_{\mu\nu}\epsilon_{\nu\rho}=\delta_{\mu\rho}$, the structure constants in this basis are given by:
	\begin{equation}
		\left[\hat{\Gamma}\left(p\right),\hat{\Gamma}\left(q\right)\right]=iF\left(p,q,-p-q\right)\hat{\Gamma}\left(p+q\right), 
		\ F\left(p,q,-p-q\right)=-\sqrt{\frac{2}{N}}\sin\left(\frac{1}{2}\theta_{\mu\nu}p_{\mu}q_{\nu}\right) \, .
	\end{equation}
	With this construction, momentum turns out to be quantised in units of $2\pi/(l_\mu l_g)$ in the twisted directions and  $2\pi/l_\mu$ in the periodic ones.
	In order to have the same momentum quantisation in all directions, we chose a torus of size $l$ in the twisted directions and $\tl=l_gl$ in the periodic ones~\cite{Keegan:2015lva}.

	We will implement a finite volume scheme similar to the one defined in~\cite{Ramos:2014kla}, introducing a flow time parameter $t$, and a new gauge field $B_\mu(x,t)$ driven by a modified flow equation:
	\begin{equation}
		\partial_{t}B_{\mu}(x,t)=D_{\nu}G_{\nu\mu}(x,t)+D_\mu \partial_\nu B_\nu(x,t),\quad B_{\mu}(x,0)=A_{\mu}(x) \, ,
	\end{equation}
	where $G_{\mu\nu}$ and $D_\mu$ are the field strength tensor and covariant derivative of the $B$ fields. The twisted gradient flow coupling is then defined as in~\cite{GarciaPerez2015}:
	\begin{equation}
		\lambda_\TGF(\tl)=Ng^2_\TGF(\tl)=\left.\mathcal{F}(c) \, \left\langle \frac{ t^2 E(t) }{N}\right\rangle \right|_{t=c^2\tl^2/8}, \quad E(t)=\half\Tr\left(G_{\mu\nu}(x,t)G_{\mu\nu}(x,t)\right)  \, ,
	\end{equation}
	where $\mathcal{F}(c)$ is a normalisation constant introduced so that $\lambda_\TGF=\lambda_0+\mathcal{O}(\lambda_0^2)$, and $c$ is a scheme-defining parameter relating the energy scale to the effective size of the torus, $\mu=1/\sqrt{8t}=1/c\tl$.

\section{LO and NLO terms in perturbation theory}

	With our coupling defined, we can start the expansion of $\langle E(t) \rangle$ around $A_\mu=0$. The general procedure is similar to the infinite volume one in \cite{Luscher:2010iy}, though the computations and regularisation procedure turn out to be quite different. We first rescale the gauge fields in the Lagrangian, $A_\mu(x) \rightarrow g_0A_\mu(x)$, and determine the Feynman rules in momentum space, which are essentially the usual ones replacing the group structure constants with $F(p,q,-p-q)$; see for instance~\cite{Perez:2014sqa} and~\cite{Perez:2017jyq} for an analogous perturbative computation on the lattice. We then expand the flowed fields in powers of the bare coupling:
	\begin{equation}
		B_{\mu}(x,t)=\underset{k}{\sum} \, g_{0}^{k}B_{\mu}^{(k)}(x,t), \qquad B_{\mu}^{\left(k\right)}\left(x,t\right)=V^{-\half}\, \sum_q^{\prime} B_{\mu}^{\left(k\right)}\left(q,t\right) \,  e^{iqx} \, \hat{\Gamma}\left(q\right) \, ,
	\end{equation}
	and bring this expansion into $E(t)$, discarding all terms beyond $\mathcal{O}(g_0^4)$. This leaves us with a sum of terms containing the expectation values of products of $B_\mu^{(k)}(p,t)$ fields, which have to be related to the original $A_\mu(p)$ fields by solving the flow equations order by order in momentum space. The solution up to third order is enough for NLO results, and so the final result is expressed as a series expansion in the bare 't Hooft coupling $\lambda_0$, parametrised as:
	\begin{equation}
		\left \langle \frac{E(t)}{N}\right \rangle \equiv\lambda_0 \, \cE^{(0)}(t)+\lambda_0^2 \, \cE^{(1)}(t)+\mathcal{O}\left(\lambda_0^3\right)\, .
	\end{equation}
	The LO term in that expansion reads:
	\begin{equation}
		\cE^{(0)}(t)=\half(d-1)\tl^{-d}\sum^{\prime}_{m\in\mathbb{Z}^d}e^{-8t\pi^2m^2/\tl^2} \equiv \frac{(d-1)}{2\,  (8\pi t) ^{d/2}}\, \, {\cA}(2 \hc t') ,
	\end{equation}
	where we introduced two auxiliary variables $\hc= \pi c^2/2$ and $t' \equiv 8 t /(c \tl)^2$ (to be set to one when computing the coupling), and a function ${\cA}$, which in terms of Jacobi theta functions $\theta_3$ is:
	\begin{equation}
		{\cA}(x) \equiv  x^{d/2} \sum^{\prime}_{m\in\mathbb{Z}^d} e^{-\pi x  m^2 }= \ \theta_3^{(d-d_t)} \left( 0 , \frac{i}{  x} \right) \left \{ \theta_3^{d_t}  \left( 0 ,\frac{i}{ x}\right) - \frac{1}{N^2} \theta_3^{d_t} \left( 0 ,\frac{i}{ x  l_g^2}\right)  \right \}
	\end{equation}
	The LO infinite volume expression corresponds to the $c\rightarrow 0, \ \tl\rightarrow \infty$ limit, with $c\tl$ kept fixed. In that limit, ${\cA}(2 \hc t')\rightarrow 1-1/N^2$ and one recovers the result from \cite{Luscher:2010iy}.

	The NLO contribution, on the other hand, can be rewritten after quite a bit of algebra in terms of twelve basic multi-dimensional integrals $I_i$ whose integrand is given in terms of the functions:
	\begin{equation}
		\Phi(s,u,v)= {\cal N}  \sum_{m,n\in\mathbb{Z}^{d}} e^{-\pi \hc (s m^2+u n^2+ 2vmn)} (1-\text{Re } e^{-2 \pi i \hat{\theta}n\tilde{\epsilon}m}), \quad \mathcal {N}= \frac{\hc ^2}{32 \pi ^2 \lef ^ {(2d-4)}} \, , \label{eq:phi}
	\end{equation}
	The integrals involve flow time integration and Schwinger parameters. For instance, one of the twelve contributing integrals reads:
	\begin{equation}
		I_{10}(\Phi, t')= \int_{0}^{\infty}dz\int_{0}^{t'}dx x\, \partial_{t'}\Phi\left(2t'+z,2t',x\right) \, .
	\end{equation}
	These $\Phi$ functions, incidentally, can be rewritten in matrix form in terms of Siegel theta functions, which are often implemented in 
	computational software such as Mathematica:
	\begin{equation}
		\Phi(s,u,v)= {\cal N}  \{\Theta(0|A( s, u, v, \htheta=0)) - \Theta(0|A( s, u, v, \htheta))\}, \ \Theta\left(z|A\right)\equiv\underset{M\in\mathbb{Z}^{2d}}{\sum}e^{i\pi\left(M^{t}AM+2z\cdot M\right)} \, ,
	\end{equation}
	with the $2d\times 2d$ matrix $A$ conveniently chosen to reproduce eq.~\eqref{eq:phi}.

\section{Identifying and regularising the divergences}	

	We have so far set up a TGF scheme to define a running coupling in perturbation theory and have expanded it up to NLO, expressing it in terms of twelve basic integrals, some of which are UV divergent. In order to regularise them we will identify the asymptotic divergence within each momentum sum and subtract it from the integrand, and then deal with the subtracted term using dimensional regularisation. The sums over momenta entering the definition of $\Phi$ can be shown to be UV convergent provided $su -v^2>0$. As the quantity $\alpha\equiv s -v^2/u$ is always positive definite within our integration ranges, the UV divergences will appear whenever $u=0$. The asymptotic divergence at $u=0$ is easily identified if one uses Poisson resummation to rewrite:
	\begin{equation}
		\sum_{m,n\in\mathbb{Z}^{d}} e^{-\pi \hc (s m^2+u n^2+ 2vmn)-2 \pi i \hat{\theta}n\tilde{\epsilon}m } = (\hc u)^{-\frac{d}{2}} \sum_{m,n\in\mathbb{Z}^{d}} e^{-\pi \hc \alpha m^2 - {\frac{\pi}{\hc u}} (n - \hat{\theta} \tilde \epsilon m)^2- i 2 \pi \frac{ v}{u} m n} 
	\end{equation}
	The divergent behaviour at $u=0$ on the right hand side occurs whenever $n=\hat{\theta}\tilde{\epsilon}m \in \mathbb{Z}^d$, and so by excluding those particular momenta from the sums the integrals are regularised. The final result for each UV-divergent integral~\footnote{Except for one of them, which requires a similar but slightly more convoluted procedure we will not detail here.} can be expressed in terms of a finite piece given by:
	\begin{equation}
		I_i^\rf (t') =  I_i(\Phi-\Phio, t') + I_i(\Phio, t')  - \cA(2\hc t') \, I_i (\Phi^\infty,t')  \, , \label{eq:Ifin}
	\end{equation}
	and a divergent one requiring analytical continuation to four dimensions:
	\begin{equation}
		I_i^\rd (t') = {\cA}(2\hc t') \, I_i (\Phi^\infty,t') \, ,
	\end{equation}
	where:
	\begin{equation}
	\Phi^{(0)}(s,u,v) = \cA\left(\hc \alpha\right)  \Phi^\infty(s,u,v) \, , \quad \Phi^\infty(s,u,v) = {\cal N}  \hc^{-d}  \left (u \alpha\right)^{-d /2}
	\, .
	\end{equation}
	$I_i(H, t')$ stands for the corresponding multiple integral, evaluated replacing the original integrand $\Phi$ with $H$. To evaluate the finite piece numerically, we used C++ and  Mathematica codes developed for the task. The divergent piece, directly given by the product of $ {\cA}(2\hc t')$ times the infinite volume result, 
was evaluated analytically in dimensional regularisation.

	\begin{figure} [t]
		\begin{minipage}{0.5\linewidth}
			\includegraphics[width=1.00\linewidth]{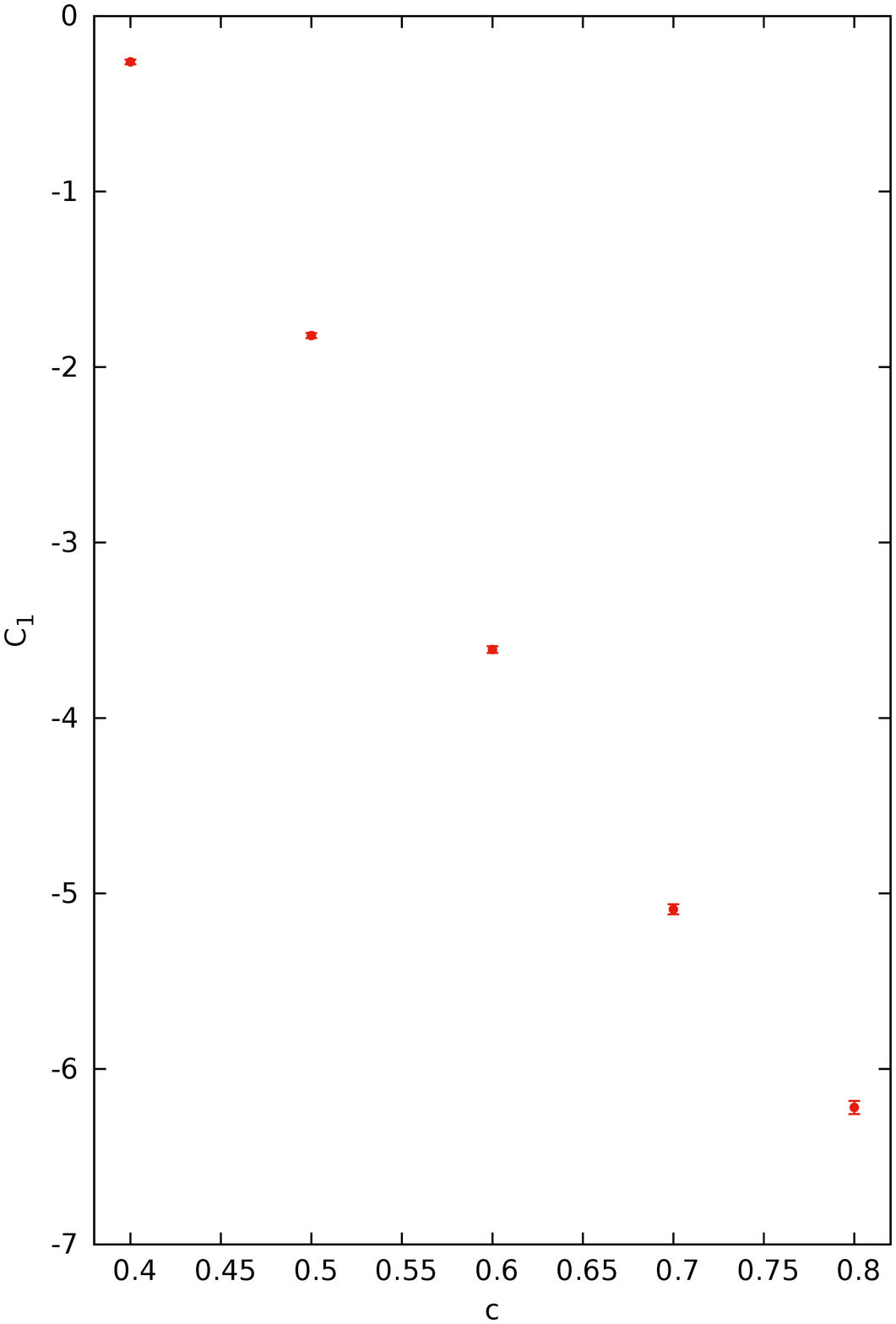}
			\caption{$\cC_1$ results for $\bar{k}=1$ in $SU(3)$}\label{fig:SU3}
		\end{minipage}
		\hfill
		\begin{minipage}{0.5\linewidth}
			\includegraphics[width=1.00\linewidth]{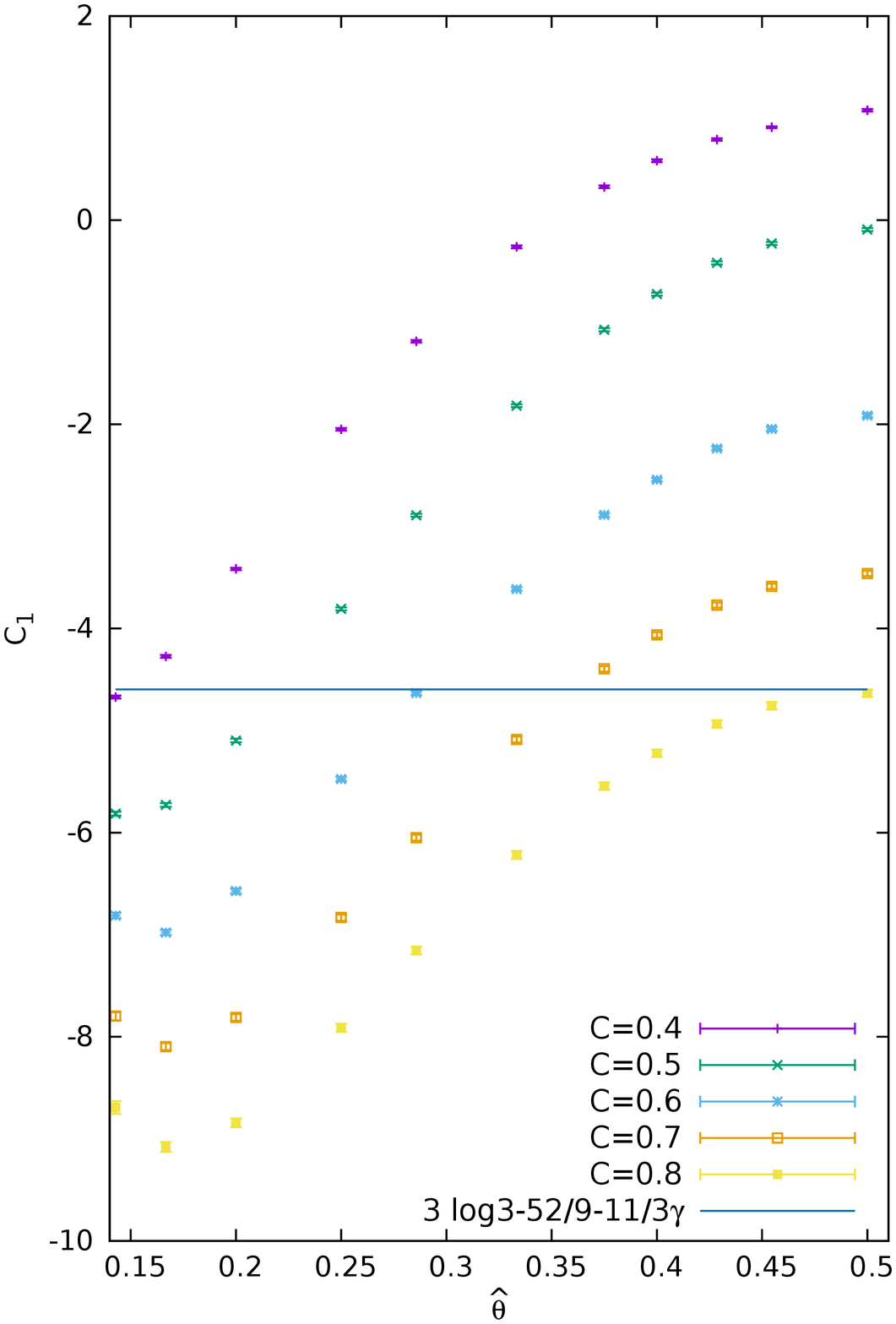}
			\caption{$\cC_1$ full results for several $\htheta= \bar{k}/N$}\label{fig:Full}
		\end{minipage}
	\end{figure}

\section{The TGF running coupling}

	With the divergences regularised, we can finally gather the results of everything done so far. Plugging the expansion of the observable into the definition of the coupling, we have:
	\begin{equation}
		\lambda_\TGF(\tl) = \lambda_0  \left \{1 + \frac{\lambda_0 ( 4 \pi c^2 \tl^{\, 2})^\epsilon}{ 16 \pi^2 } \left ( \frac{11}{3\epsilon} + \frac{52}{9} -3 \log 3 + \cC_1 \right)  \right \} \, ,
	\end{equation}
	where $\cC_1$ denotes the sum of all finite contributions we separated in the previous section (with their corresponding coefficients), evaluated at $t'=1$. 	The infinite volume result from \cite{Luscher:2010iy} corresponds to $\cC_1=0$ and is correctly reproduced by our formula. The bare coupling constant can be replaced by its expression at 1-loop in the $\MS$ scheme to give:
	\begin{equation}
		\lambda_\TGF(\tl)=\lambda_\MS(\mu)\left\{1+\frac{\lambda_\MS(\mu)}{16\pi^2}\left(\frac{11}{3}\gamma_E+\frac{52}{9}-3\log 3 + \cC_1 \right) \right \}, 
	\end{equation}
	with the $\MS$ scale $\mu= 1/(c \tl)$. As for the ratio between $\Lambda$ parameters in both schemes:
	\begin{equation}
		\log \left (\frac{\Lambda_\TGF}{\Lambda_\MS} \right) = \frac{1}{32\pi^2 b_0} \left (\frac{11}{3} \gamma_E + \frac{52}{9} -3 \log 3 + \cC_1\right)  \, .
	\end{equation}

	Using the integration codes mentioned in the previous section, we computed $\cC_1$ for a range of values of $c$, $\bar{k}$ and $N$. The results for $SU(3)$ with magnetic flux $\bar{k}=1$ at various  values of $c$ are presented on fig.\ref{fig:SU3}~\footnote{The present plots correct an error found in the plots shown in the presentation slides.}. Figure~\ref{fig:Full} displays the value of $\cC_1$ as a function of $\hat{\theta}= \bar k /l_g$, obtained for several gauge groups ($N=2,3,4,5,6,7)$ with magnetic flux $\bar k=1$, as well as for the sets $(\bar k,N)= (2,5), (2,7), (3,7), (3,8), (5,11)$. A full list of results including smaller values of $c$ and an analysis of the large $N$ and infinite volume limits will be presented in \cite{us}.

\section{Summary and future prospects}

	We have considered a pure gauge $SU(N)$ theory in the continuum, defined on a finite volume, four-dimensional torus with twisted boundary conditions on one plane and periodic boundary conditions in the rest. The torus was chosen to have a period of $l$ in the twisted directions and $\tilde{l}=Nl$ in the rest, so as to have all momenta quantised in terms of a single scale. We used the gradient flow to define a renormalised running 't Hooft coupling from the action density of the theory, using $\tilde{l}$ as the energy scale for the running. We determined the expressions of both the LO and the NLO perturbative contributions in terms of quantities that can be computed numerically. The NLO was parametrised in terms of twelve basic integrals, some of which were divergent. We devised a way to regularise such integrals by isolating the divergent terms and relating them to infinite volume integrals treatable in dimensional regularisation. We then numerically computed the finite part of these integrals for several $SU(N)$ gauge groups and a range of values of the twist related factor $\hat{\theta}=\bar k/N$ and $c$, and presented the results.

	In the future, we plan to repeat this computation non-perturbatively on the lattice, defining a running coupling in terms of $\tilde{l}$ and computing its running for $SU(3)$ using step scaling techniques. The inclusion of adjoint fermions, and an estimation of the extent of finite volume corrections in relation to the matter of volume independence conjectures would also constitute interesting continuations of this work.

\section*{Acknowledgements}

	We would like to thank Antonio Gonz\'alez-Arroyo and Alberto Ramos for many valuable discussions on both this topic and related ones. We acknowledge financial support from the MINECO/FEDER grant FPA2015-68541-P  and the MINECO Centro de Excelencia Severo Ochoa Programs SEV-2012-0249 and SEV-2016-0597. E.I. Bribi\'an acknowledges support under the FPI grant BES-2015-071791. The numerical computations have been carried out at the IFT Hydra cluster and with computer resources provided by CESGA (Galicia Supercomputing Centre).

\providecommand{\href}[2]{#2}\begingroup\raggedright\endgroup

\end{document}